# Estimation of Condition Number of Quasi-static Darwin Model

Shingo Hiruma, *Member, IEEE,* Takeshi Mifune, *Member, IEEE,* and Tetsuji Matsuo, *Member, IEEE*

Graduate School of Engineering, Kyoto University, Kyoto, 615-8510, Japan

**This study discusses the estimation of the condition number of the quasi-static Darwin model, which is shown to be a nearly singular system of equations with a large kernel derived from the gauge invariance of full Maxwell's equations. Inequality evaluations for a minimum nonzero singular value, maximum singular value, and condition number for a nearly singular system of equations and its augmented system are provided. This discussion reveals the nearly singular nature of the Darwin model and proves the effectiveness of the augmented system method for solving the Darwin model.**

***Index Terms*—Nearly singular system of equations, Darwin model, condition number, and implicit error correction/singular decomposition technique.**

## I. Introduction

THE Darwin model [1]-[4] is gaining attention because it can treat inductive, resistive, and capacitive effects simultaneously in the time-domain analyses. This feature is promising for designing electromagnetic devices for high-frequency applications. However, the Darwin model is known to be a severely ill-conditioned problem [1][2] and is difficult to solve using conventional Krylov subspace solvers. Recently, a stabilized version of the Darwin model was proposed in [4][5], but the mechanism for convergence improvement was not discussed. This study aims to quantitatively evaluate the ill-conditioning of the Darwin model and explain the mechanism for the convergence improvement of the stabilized version.

First, we focus on the nature of nearly singular systems of equations [6], which arise in many examples, such as the finite element discretization for $H(\text{grad})$, $H(\text{div})$ and $H(\text{curl})$ systems including eddy current problems. We provide an elementary estimation of the condition number by evaluating the nonzero minimum and maximum singular values of the nearly singular and augmented systems using the augmented system method [6], which is also referred to as the implicit error correction/singular decomposition technique (IEC/SDT) [7][8].

It is then shown that the Darwin model falls into a class of nearly singular systems, which has a near-kernel derived from the gauge invariance of full Maxwell's equations. We prove that the Darwin model contains considerably small singular values that depend on the frequency and permittivity, which makes severely ill-conditioned. In addition, we prove that frequency-stable convergence can be achieved using the augmented system method, leading to stabilized version of the Darwin model.

## II. Formulation

The Darwin approximation expresses the displacement current in Ampère's law using the longitudinal wave component of the electric field $\boldsymbol{E}_L$, neglecting the transverse wave component $\boldsymbol{E}_T$. Ampère-Darwin's law is given by

$$\nabla \times \boldsymbol{H} = \boldsymbol{J}_s + \sigma \boldsymbol{E} + j\omega\varepsilon \boldsymbol{E}_L \tag{1}$$

where $\boldsymbol{H}, \boldsymbol{J}_s, \boldsymbol{E}, \omega, \sigma, \varepsilon, j$ denote the magnetic field, solenoidal source current density, electric field, angular frequency, conductivity, permittivity, and imaginary unit, respectively. Using the magnetic vector potential $\boldsymbol{A}$ and electric scalar potential $\varphi$, we can write $\boldsymbol{E} = -j\omega\boldsymbol{A} - \nabla\varphi$, $\boldsymbol{E}_L = -\nabla\varphi$. Considering the Gauss's law for the transverse wave component of the displacement current $\boldsymbol{D}_T = -j\omega\varepsilon\boldsymbol{A}$, the Coulomb-type gauge condition is derived as follows

$$\nabla \cdot \boldsymbol{D}_T = -j\omega\nabla \cdot (\varepsilon\boldsymbol{A}) = 0 \tag{2}$$

which is necessary to obtain a unique solution to the Darwin model. Discretizing (1) and (2) using the edge-based finite element method with appropriate boundary conditions, we obtain

$$\begin{bmatrix} C^\top[\nu]C + j\omega[\sigma] & [\sigma]G + j\omega[\varepsilon]G \\ j\omega G^\top[\varepsilon] & 0 \end{bmatrix} \begin{bmatrix} a \\ \phi \end{bmatrix} = \begin{bmatrix} b \\ 0 \end{bmatrix} \tag{3}$$

where $G$ and $C$ are the discrete gradient and curl matrices, respectively, satisfying $CG = 0$, $[\nu], [\sigma]$ and $[\epsilon]$ are the reluctivity, conductivity and permittivity matrices, respectively, $a$ and $\phi$ are the unknown vectors for the magnetic vector and electric scalar potentials, and $b$ is the right-handed side vector satisfying $G^\top b = 0$. This equation is a generalized saddle-point problem, and the main difficulty in solving the Darwin model arises from the nature of the generalized saddle-point problem: 1) the matrix is typically indefinite, which can lead to numerical instability in iterative solvers; 2) the eigenvalues can vary over a wide range, resulting in a large condition number and a slow convergence of iterative methods; and 3) scaling is challenging because of significant differences between coefficients of matrix blocks, which is caused by the high contrast in material coefficients such as conductivity and permittivity [1][2].

In this study, we analyze the condition number of the Darwin model from the viewpoint of nearly singular systems. We also explain that the augmented system method is an efficient way to improve the convergence of the iterative methods. To do so, we focus on the symmetrized formulation of the Darwin model [4]. By multiplying both sides of the first row of (3) by $(j\omega)^{-1}G^{\top}$ from the left and adding it to the second row, it can be symmetrized as

$$\begin{bmatrix} C^{\top}[\nu]C + j\omega[\sigma] & [\sigma]G + j\omega[\varepsilon]G \\ G^{\top}[\sigma] + j\omega G^{\top}[\varepsilon] & \frac{1}{j\omega}G^{\top}[\sigma]G + G^{\top}[\varepsilon]G \end{bmatrix} \begin{bmatrix} a \\ \phi \end{bmatrix} = \begin{bmatrix} b \\ 0 \end{bmatrix}. \quad (4)$$

Although we remain in the frequency-domain, the discussion is also valid for the Darwin model in the time-domain.

## III. Methodology

### A. Nearly Singular Systems of Equations

We consider the following equation

$$(A_0 + \epsilon A_1)x = b \quad (5)$$

where $A_0$ is a (real or complex) $n$-dimensional matrix with a kernel space $\mathrm{Ker}(A_0)$, $A_1$ is a perturbation matrix, and $\epsilon$ is a small parameter. Systems of this form are referred to as *nearly singular systems of equations* [6], which arise in many applications. When the parameter $\epsilon$ is very small, the component of the solution belonging to $\mathrm{Ker}(A_0)$ becomes very small under the A-inner product, making it difficult to approximate the solution and deteriorating the convergence of Krylov subspace solvers.

In the next subsection, we derive inequalities that provide the upper bounds for the nonzero minimum singular values, lower bounds for the maximum singular values, and lower bounds for the condition numbers of the coefficient matrix in (4). The proposed method enables the quantitative evaluation of the condition number, allowing for an explanation of the poor convergence behavior of the preconditioned conjugate gradient (CG) methods.

### B. Estimation of condition number

First, let $B$ be an $m$-dimensional matrix representing the basis vectors of $\mathrm{Ker}(A_0)$. Matrix $B$ may depend on $\varepsilon$, and we can write it as $B(\epsilon)$. Then, the following relationships hold

$$A_0 B = 0, \qquad AB = \epsilon A_1 B. \quad (6)$$

where $A = A_0 + \epsilon A_1$. The nonzero minimum singular value $\sigma_{\min}^0$ of the coefficient matrix $A$ can be calculated using the nonzero minimum eigenvalue $\mu_{\min}$ of the Hermitian matrix $A^*A$, where $A^*$ is the conjugate transpose of $A$, as follows [9]

$$(\sigma_{\min}^0)^2 = \mu_{\min} = \min_{\substack{x \in \mathrm{Ran}(A^*) \\ x \neq 0}} \frac{\|Ax\|^2}{\|x\|^2} \quad (7)$$

where $\mathrm{Ran}(A^*)$ is a column space of $A$. An upper bound for the nonzero minimum singular value of the coefficient matrix of a nearly singular system of equations can be obtained by substituting the basis vectors of $\mathrm{Ker}(A_0)$ into (7). Specifically, it can be expressed as follows

$$\mu_{\min} \leq \min_{By \neq 0} \frac{\|ABy\|^2}{\|By\|^2} = \min_{By \neq 0} \frac{\epsilon^2\|A_1By\|^2}{\|By\|^2} = C_0(\epsilon)\epsilon^2 \quad (8)$$

where $C_0(\epsilon)$ is the nonzero minimum eigenvalue of the following generalized eigenvalue problem: $B^*A_1^*A_1By = \lambda(\epsilon)B^*By$. If matrix $B$ depends on $\epsilon$, the eigenvalues of the generalized eigenvalue problem also depend on $\epsilon$. From the above, the nonzero minimum singular value can be estimated as follows

$$\sigma_{\min}^0 \leq \sqrt{C_0(\epsilon)}\,\epsilon. \quad (9)$$

Equation (9) explicitly expresses the $\epsilon$-dependence of the nonzero minimum singular value. Furthermore, the lower bound of the maximum singular value is given by the Frobenius norm of the matrix $A$

$$(\sigma_{\max})^2 \geq \frac{1}{n}\|A\|_{\mathrm{F}}^2 = \frac{1}{n}\sum_{0 \leq i,j < n} |a_{ij}|^2 = C_1. \quad (10)$$

where $C_1$ is a constant number. Therefore, the lower bound for the condition number is

$$\kappa = \frac{\sigma_{\max}}{\sigma_{\min}^0} \geq \sqrt{\frac{C_1}{C_0}}\frac{1}{\epsilon} \to \infty, \qquad \text{when } \epsilon \to 0. \quad (11)$$

It can be observed that the condition number diverges as $\epsilon$ decreases. When $\kappa$ is large, the number of iterations in the iterative method such as the CG method increases significantly. Note that ordinary preconditioning techniques, such as incomplete Cholesky (IC) decomposition, are less effective for nearly singular systems in reducing the number of iterations.

### C. Augmented system method

The augmented system method, which introduces auxiliary variables into the equations, is known to be an effective solution for improving the convergence properties of nearly singular systems.

In [10], the behavior of the minimum eigenvalue of the coefficient matrix of the augmented system when $\epsilon \to 0$ was discussed based on the continuity of the eigenvalues. In addition, the convergence rates of stationary iterative methods, such as the Gauss-Seidel and Jacobi iterations, were discussed in [5][11]. Here, we aim to derive an inequality for the lower bound of the nonzero minimum singular value of the coefficient matrix of the augmented system to demonstrate that parameter-independent convergence is achieved in the augmented system.

For (5), the augmented system is defined as follows

$$A'x' = \begin{bmatrix} A & \epsilon A_1 B \\ \epsilon B^{\top} A_1 & \epsilon B^{\top} A_1 B \end{bmatrix} \begin{bmatrix} x \\ p \end{bmatrix} = \begin{bmatrix} b \\ B^{\top} b \end{bmatrix}. \quad (12)$$

The augmented system is an underdetermined system because the first and second rows are linearly dependent. Therefore, for any $\chi$, the vector obtained by the following transformation is also a solution of the augmented system

$$\begin{bmatrix} x' \\ p' \end{bmatrix} = \begin{bmatrix} x \\ p \end{bmatrix} - \begin{bmatrix} -B\chi \\ \chi \end{bmatrix} \tag{13}$$

which means that $A'$ is rank deficient and has zero eigenvalues. The number of zero eigenvalues is equal to the dimension of $B$. The kernel and range spaces of $A'$ are explicitly given as follows

$$\mathrm{Ker}(A') = \left\{ \begin{bmatrix} -B\chi \\ \chi \end{bmatrix} \middle| \chi \in \mathbb{C}^m \right\}, \mathrm{Ran}(A') = \left\{ \begin{bmatrix} v \\ B^\top v \end{bmatrix} \middle| v \in \mathbb{C}^n \right\}. \tag{14}$$

Without preconditioning, as $\epsilon \to 0$, the nonzero minimum singular value of $A'$ approaches zero, as can be understood from the analysis of eigenvalue continuity [10]. Thus, we consider the diagonal preconditioning to obtain the lower bound. We first define the diagonal matrix of $A'$ as

$$D_{A'} = \begin{bmatrix} D_A & 0 \\ 0 & D_{B^\top AB} \end{bmatrix} = \begin{bmatrix} D_A & 0 \\ 0 & \epsilon D_{B^\top A_1 B} \end{bmatrix} \tag{15}$$

where $D_A$, $D_{B^\top AB}$ and $D_{B^\top A_1 B}$ are the diagonal matrices of $A$, $B^T AB$ and $B^\top A_1 B$, respectively. The nonzero minimum singular value $\sigma_{\min}^{a0}$ of $A'$ is defined as

$$(\sigma_{\min}^{a0})^2 = \lambda_{\min} = \min_{\substack{x \in \mathrm{Ran}(A'^*) \\ x \neq 0}} \frac{\left\| D_{A'}^{-1} A' x \right\|^2}{\|x\|^2}. \tag{16}$$

By substituting $x = [v^\top \quad (B^* v)^\top]^\top \in \mathrm{Ker}(A')^\perp = \mathrm{Ran}(A'^*)$ into (16), we obtain

$$\sigma_{\min}^{a0} \geq \sqrt{\min_{v \neq 0} \frac{\left\| D_{A_0}^{-1} A_0 v \right\|^2 + \left\| D_{B^\top A_1 B}^{-1} B^\top A_1 v \right\|^2}{\|v\|^2}} \tag{17}$$

when $\epsilon \to 0$. This means that the nonzero minimum singular value of the diagonally scaled matrix is bounded below by a constant value that depends on $D_{A_0}^{-1} A_0$ and $D_{\mathrm{B}^\top A_1 B}^{-1} B^\top A_1$, and does not approach zero as $\epsilon \to 0$. The maximum singular value also hardly depends on $\epsilon$ when it is small and $A_0$ is independent of $\epsilon$, and therefore, the condition number is also almost independent of $\epsilon$. Consequently, this inequality demonstrates parameter-independent convergence of the augmented system.

## IV. Application to Darwin Model

### A. Condition Number Estimation of Darwin Model

Because the Darwin model neglects the transverse wave component of the displacement current, the equation does not exhibit gauge invariance. This can be explicitly expressed by writing the symmetrized Darwin formulation (4) as

$$(A_0 + \omega^2 \varepsilon_0 A_1) \begin{bmatrix} a \\ \phi \end{bmatrix} = \begin{bmatrix} b \\ 0 \end{bmatrix}, \tag{18}$$

$$A_0 = \begin{bmatrix} C^\top [\nu] C + j\omega[\sigma] + (j\omega)^2 [\varepsilon] & [\sigma] G + j\omega [\varepsilon] G \\ G^\top [\sigma] + j\omega G^\top [\varepsilon] & \frac{1}{j\omega} G^\top [\sigma] G + G^\top [\varepsilon] G \end{bmatrix}, \tag{19}$$

$$A_1 = \begin{bmatrix} [\varepsilon_r] & 0 \\ 0 & 0 \end{bmatrix} \tag{20}$$

where $A_0$ is the finite element matrix obtained by formulating full Maxwell's equations using the A-V formulation. As is well known, Maxwell's equations have gauge invariance, and the magnetic vector and electric scalar potentials are indeterminate. Reflecting this, $A_0$ has $m$ zero eigenvalues, and the corresponding zero eigenvectors can be written as

$$A_0 \begin{bmatrix} -G \\ j\omega I \end{bmatrix} \phi = 0, \qquad \forall \phi \in \mathbb{C}^m \tag{21}$$

where $I$ denotes an $m$-dimensional identity matrix. Additionally, by setting $B = [-G^\top \quad j\omega I]^\top$, we obtain $A_0 B = 0$. Therefore, the Darwin model can be regarded as a nearly singular system of equations with a coefficient matrix containing a matrix with a kernel space arising from the gauge invariance of full Maxwell's equations. It should be emphasized that in the eddy current problem the kernel space is derived from the algebraic relationship between the discrete curl and gradient matrices $CG = 0$, whereas in the Darwin model it is derived from the gauge invariance.

Based on the discussion in Sec. III-B, the nonzero minimum singular value of the coefficient matrix of linear system (18) can be estimated as follows

$$\sigma_{\min}^{0} \leq \sqrt{\frac{C_0}{\lambda_m + \omega^2}} \omega^2 \varepsilon_0 \tag{22}$$

where $\lambda_m$ is the minimum eigenvalue of the discrete Laplacian $G^\top G$, and $C_0$ is the minimum eigenvalue of the matrix $G^\top [\varepsilon_r][\varepsilon_r] G$.

The maximum singular value, $\sigma_{\max}$, is influenced by $\omega$ and $\varepsilon$ due to the dependence of $A$ on these parameters. However, this is not critical because it can be effectively mitigated through the use of preconditioning techniques. More importantly, we establish a constant lower bound for $\sigma_{\max}$, as shown in (10), which is sufficient for our analysis. Therefore, the estimation of the condition number becomes

$$\kappa \geq \sqrt{\frac{\lambda_m + \omega^2}{C_0}} \frac{\sqrt{C_1}}{\omega^2 \varepsilon_0}. \tag{23}$$

As $\omega \to 0$, the nonzero minimum singular value approaches zero, causing the condition number to diverge to infinite. At high frequencies, the lower bound becomes smaller, but the condition number remains large due to the presence of $\varepsilon_0^{-1}$ where $\varepsilon_0$ is on the order of $10^{-12}$ in SI units which is a considerably small value. This bound clearly explains why the Darwin model is severely ill-conditioned.

### B. Augmented system of Darwin model

We obtain the augmented system (12) of the Darwin model using the basis of the kernel space $B = [-G^\top \quad j\omega I]^\top$ [4]. From (17), the lower bound for the nonzero minimum singular value of the coefficient matrix of the augmented is given as

$$\sigma_{\min}^{a0} \geq \sqrt{\min_{v \neq 0} \frac{\left\|D_{A_0}^{-1} A_0 v\right\|^2 + \left\|\left[D_{G^\top[\varepsilon_r]G}^{-1} G^\top[\varepsilon_r] \quad 0\right] v\right\|^2}{\|v\|^2}} \quad (24)$$

when $\omega^2 \varepsilon_0$ is small. The lower bound does not approach zero as $\omega \to 0$. Although, the condition number still depends on $\omega$ due to the presence of $\omega$ in $A_0$, the singularity of the condition number is removed in the augmented system. Therefore, this estimation justifies the use of the augmented system of the Darwin model to achieve a stable and fast convergence of the Krylov subspace solvers.

## V. Numerical Analysis

To verify inequality (22), we considered the conductor model shown in Fig. 1. Owing to the symmetry of the model, we considered a quarter-domain. A conductor with $\sigma = 10^7$ S/m, $\mu_r = 1, \varepsilon_r = 1$ was placed at the center, surrounded by air with $\sigma = 0, \mu_r = 1, \varepsilon_r = 1$.

We discretized the Darwin model using the edge-based finite element method, applied diagonal scaling preconditioning and calculated the frequency dependence of the nonzero minimum and maximum singular values of the coefficient matrix of the linear system (4). In this calculation, we set to $\varepsilon_r = 10^4$ over the entire model to scale the singular value and to avoid round-off error. The results are shown in Fig. 2a, where asymptotic behavior is also indicated. These plots are consistent with the behavior predicted by (22). Fig. 2b shows the dependence of the singular values on the relative permittivity. The latter was artificially varied over the entire model, and the frequency was set to $10^9$ Hz. Again, the numerical results are consistent with (22) because the $\sqrt{C_0}$ is proportional to $\varepsilon_r$.

To confirm the improvement in the convergence of the Darwin model using the augmented system method, we solved the original equation (4) and its augmented system (12) for $10^7$ Hz with IC preconditioned conjugate orthogonal CG (IC-COCG), as shown in Fig. 3a. We can see that the augmented system converges to a relative residual norm of $10^{-15}$ whereas the original equation exhibits oscillatory behavior in the residual, failing to converge even after 2000 iterations. In addition, we solved the Darwin model using the augmented system for $10^1$ Hz to $10^9$ Hz, as shown in Fig. 3b. We can see that the convergence property rarely depends on frequency. This confirms the realization of a frequency-stable convergence property using the augmented system method.

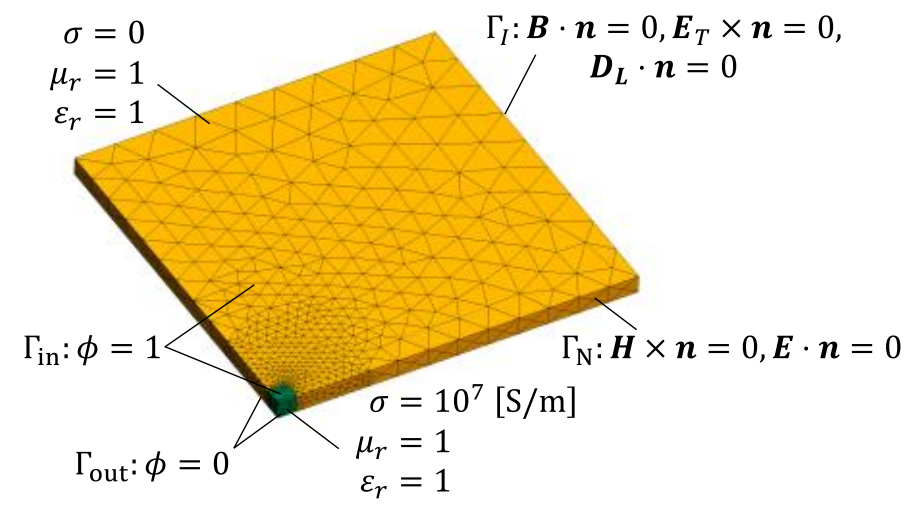

Fig. 1 Conductor model used to validate the proposed inequality.

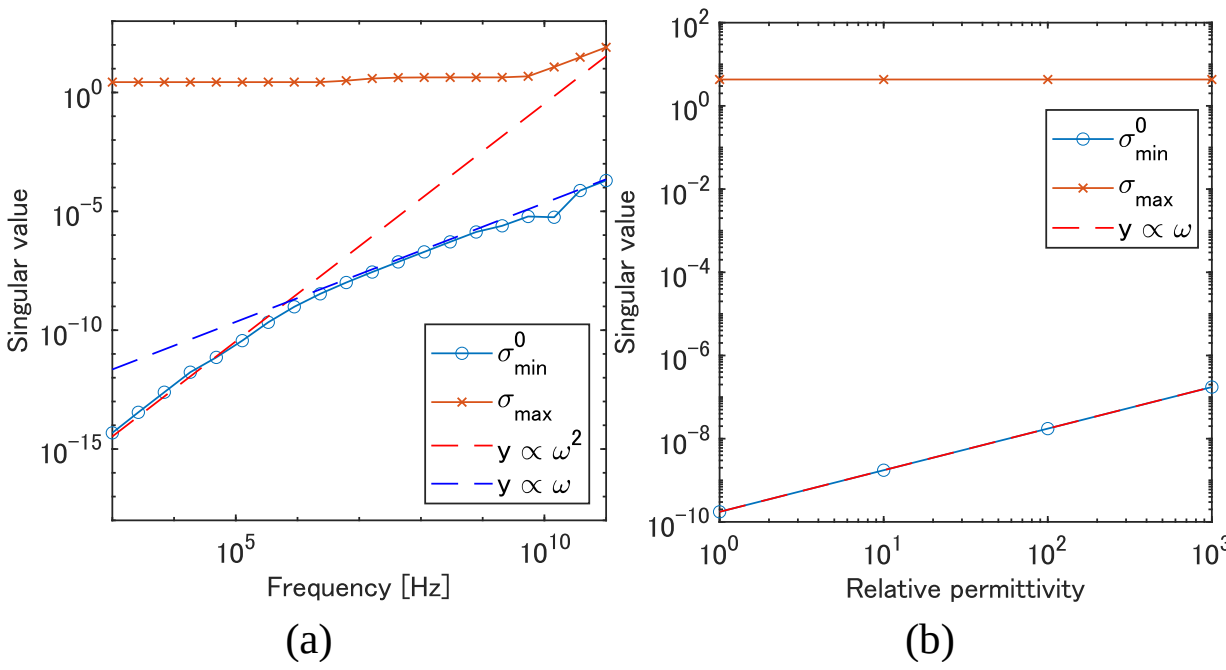

Fig. 2 Dependence of the nonzero minimum and maximum singular values on (a) frequency and (b) relative permittivity.

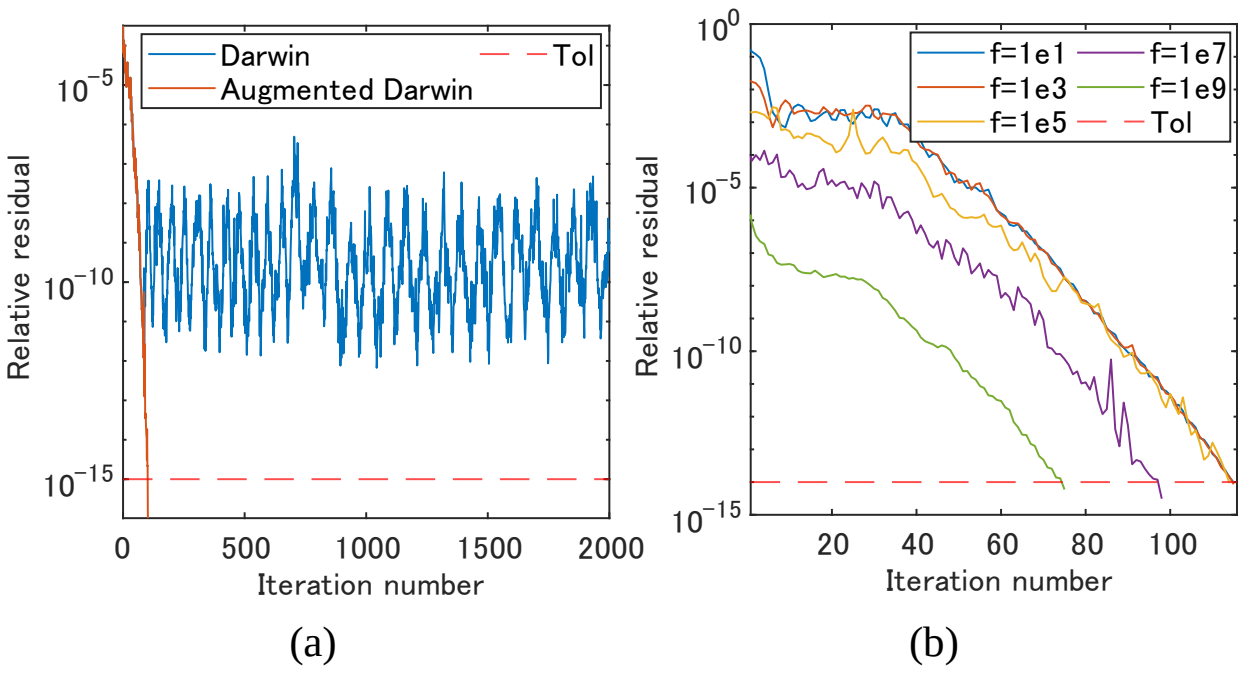

Fig. 3 (a) Convergence property of the Darwin model and its augmented system and (b) frequency-stable convergence of the Darwin model using augmented system method.

## Acknowledgment

This work was supported in part by the JSPS KAKENHI Grant Numbers 22K14237 and 24K17261.